\documentclass[]{raa}            
\usepackage{graphicx,times}
\usepackage{natbib}

\def\etal{et al.}
\def\cstar{CSTAR }
\def\ssd{\ensuremath{^\circ\hspace{-0.09em}\mathrm{C}}\,}

\begin{document}

   \title{Testing and Data Reduction of the Chinese Small Telescope Array (\cstar) for Dome A, Antarctica}

   \volnopage{ {\bf 2009} Vol.\ {\bf 9} No. {\bf XX}, 000--000}
   \setcounter{page}{1}

   \author{Xu ZHOU\inst{1,4}, Zhenyu WU\inst{1}, Zhaoji
     JIANG\inst{1,4},   Xiangqun CUI\inst{2,4},  Longlong
     FENG\inst{3,4},  Xuefei Gong\inst{2,4}, Jingyao
     HU\inst{1,4},Qisheng LI\inst{1},  Genrong LIU\inst{2}, Jun
     MA\inst{1}, Jiali WANG\inst{1,4}, Lifan WANG\inst{3,4},
     Jianghua WU\inst{1}, Lirong XIA\inst{2}, Jun YAN\inst{1,4},
     Xiangyan YUAN\inst{2,4},    Fengxiang ZHAI\inst{2}, Ru
     ZHANG\inst{2}, Zhenxi ZHU\inst{3,4}}


   \institute{National Astronomical Observatories, Chinese Academy of
     Sciences, Beijing 100012, China; {\it zhouxu@bao.ac.cn}\\
    \and National Astronomical Observatories/Nanjing Institute of
    Astronomical Optics \& Technology\\ \and Purple Mountain
    Observatory\\ \and Chinese Center for Antarctic Astronomy\\ \vs
    \no  \small Received [year] [month] [day]; accepted [year] [month]
         [day] }

   \abstract{The Chinese Small Telescope ARray (hereinafter CSTAR) is
     the first Chinese astronomical   instrument on the Antarctic ice
     cap. The low temperature and low pressure testing of the data
     acquisition system was carried out in a laboratory refrigerator
     and on the 4500m Pamirs high plateau, respectively. The results
     from the final four nights of test observations demonstrated that
     \cstar was ready for operation at Dome A, Antarctica. In this
     paper we present a description of \cstar and the performance
     derived from the test observations.
   \keywords{instrumentation: detectors --- techniques: photometric --- stars: variables}
}

   \authorrunning{X. ZHOU, Z. Y. Wu, Z. J. Jiang, \etal } 
   \titlerunning{\cstar: testing and data reduction}  
   \maketitle

\section{Introduction}

Site testing at the South Pole (90\dg south, 2835\,m elevation) and
Dome C (123\dg east, 75\dg south, 3260\,m elevation) over the past
decade has shown that the Antarctic plateau offers outstanding sites
for astronomical observations. The extremely cold temperatures lead to
very low infrared backgrounds and atmospheric water vapor content. The
very low wind speeds and stable middle and upper atmosphere result in
favorable seeing conditions for high-resolution imaging
\citep{sto07}. The median free-atmosphere seeing in Dome C is 0.27
arcsec, and it is below 0.15 arcsec for 25 per cent of the time
\citep{la04}. In addition, the long dark winter on the Antarctic
plateau allows continuous observations of variable astronomical
objects.

Dome A (77\dg 21\arcmin east, 80\dg 22\arcmin south, 4093\,m
elevation), the highest point on the Antarctic plateau, is widely
predicted to be an even better astronomical site even than Dome C,
based on the topographic similarity and Dome A's higher altitude. In
January 2005, via overland traverse, Dome A was first visited by the
Polar Research Institute of China (hereinafter PRIC). This provides
astronomers with a good opportunity to explore this special area for
astronomy. PRIC plans to establish a permanently manned station at
Dome A within the next decade, with astronomy as one of the scientific
goals of the station. As part of this program, PRIC conducted a second
expedition to Dome A, arriving via overland traverse in January
2008. On this expedition, the first Chinese Antarctic astronomical
instrument, CSTAR, was deployed to Dome A. Beside the task of the
astronomical site testing, the main scientific goals of \cstar include
variable star light curves and statistics, supernovae studies,
gamma-ray burst optical afterglow detection and exoplanet detection.

\cstar was designed and constructed during 2006 -- 2007 at the
National Astronomical Observatories of China (NAOC) and the Nanjing
Institute of Astronomical Optics Technology of China (NIAOT). A series
of tests were performed on \cstar before the second expedition to Dome
A to enure that it was ready for deployment. As a result of this
careful preparation, CSTAR operated successfully during 2008, as part
of the Plateau Observatory (PLATO) at Dome A  \citep{yang09}. In
Section \ref{ins}, we describe the design and construction of
\cstar. Test observations at the Xinglong station of NAOC and the data
reduction are presented in Section \ref{obs}. Finally, a summary is
given in Section 4.

\section{instruments}\label{ins}
\begin{figure}
\includegraphics[width=100mm,angle=90]{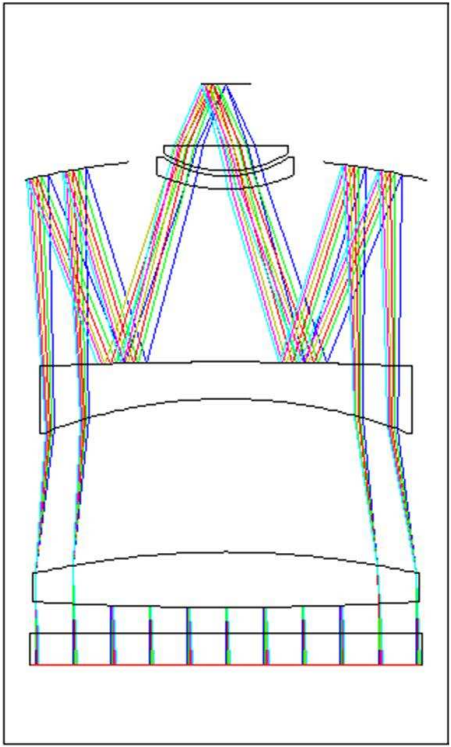}
\caption{Optical design of \cstar telescope.}\label{fig1}
\end{figure}

\begin{figure}
\includegraphics[width=80mm]{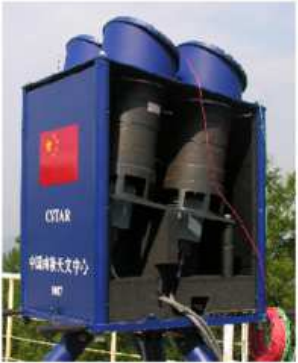}
\caption{A picture of \cstar enclosure was taken in XingLong station of NAOC.}\label{fig2}
\end{figure}

\cstar is a small $2\times2$ Schmidt-Cassegrain telescope array.  Each
telescope of \cstar has an entrance pupil diameter of 145\,mm
(effective aperture of 100\,mm) and a focal ratio of f/1.2, giving a
field of view of $\sim 4.5\dg\times4.5\dg$ . Fig.\,\ref{fig1} shows
the optical design of the \cstar telescope, which consists of a
catadioptric objective with spherical primary mirror, delivering low
chromatic aberration. The first plano lens serves both as a window and
as a filter. In order to keep the focus unchanged through a $\sim
100\ssd$ temperature difference (from 20 to $-80\ssd$), Zerodur and
fused silica are used for the main optical components and Invar 36 is
used for the telescope tube. The tube is designed to be light weight,
well sealed, and easy to assemble. The inside of the telescope tube
was filled by pure nitrogen to avoid ice and frost formation on the
internal optical surfaces. Each telescope tube is hermetically sealed
and an ITO (Indium-Tin-Oxide) film was coated onto the front
window. An electric current is passed through this film, providing
$\sim 10$ W of power to keep the surface of the window warmer than
ambient. \cstar is specifically designed for Antarctic operation,
having no moving parts at all---including the optics and mechanical
supporting system. The four telescopes are installed in a steel
enclosure, see Fig.\,\ref{fig2}, and are pointed at the South
Celestial Pole; ie., each telescope is inclined $9\dg38'$ from the
zenith. Details of the CSTAR telescope structure are described in
Yuan et al. (2008).

\begin{figure}
\includegraphics[width=100mm,angle=-90]{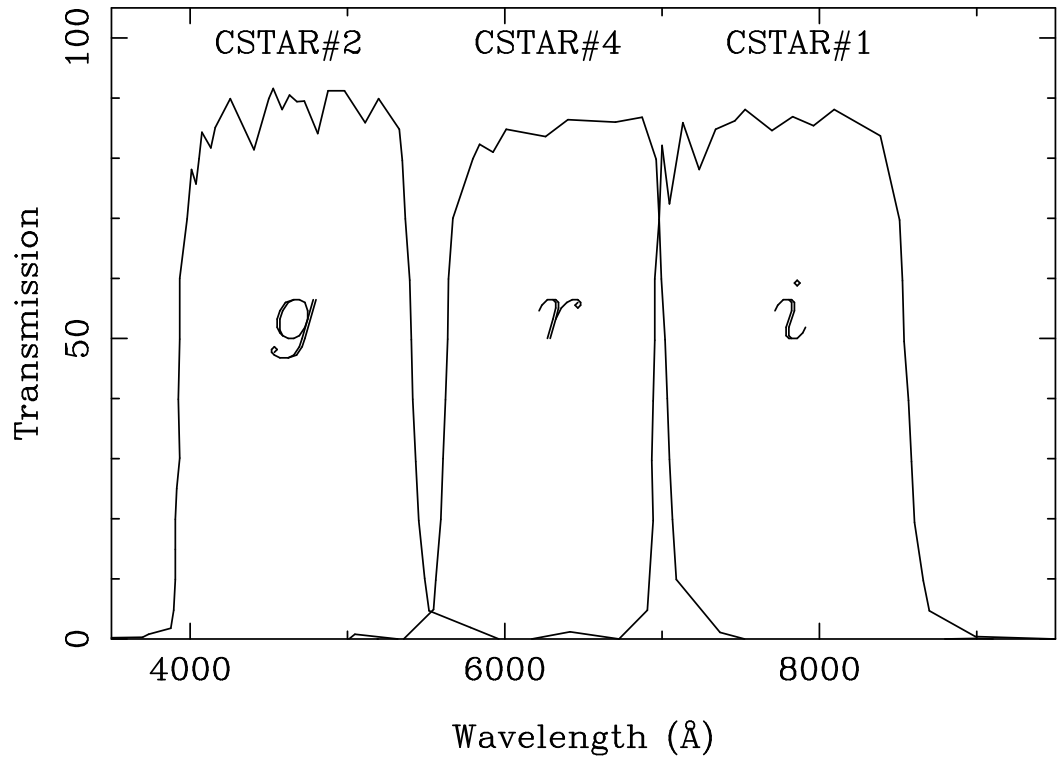}
\caption{Transmission profiles of the 3 \cstar filters. The filter codes
(see Table\,\ref{tb1} are labeled on each filter. Note that  \cstar\,\#3 has no filter.}\label{fig3}
\end{figure}

The three telescopes \cstar$\#2$, $\#4$, and $\#1$ have fixed filters:
$g$, $r$, and $i$, the fourth telescope \cstar$\#3$ is
filter-less. The main parameters of the three filters are listed in
Table \ref{tb1} and the transmission curves of those filters are
presented in Fig.\,\ref{fig3}. The filters are designed to be similar
to the corresponding filters of the SDSS \citep{fu96}. Using these
filters, \cstar can obtain multicolor photometric data for each object
simultaneously.

\begin{figure}
\includegraphics[width=125mm]{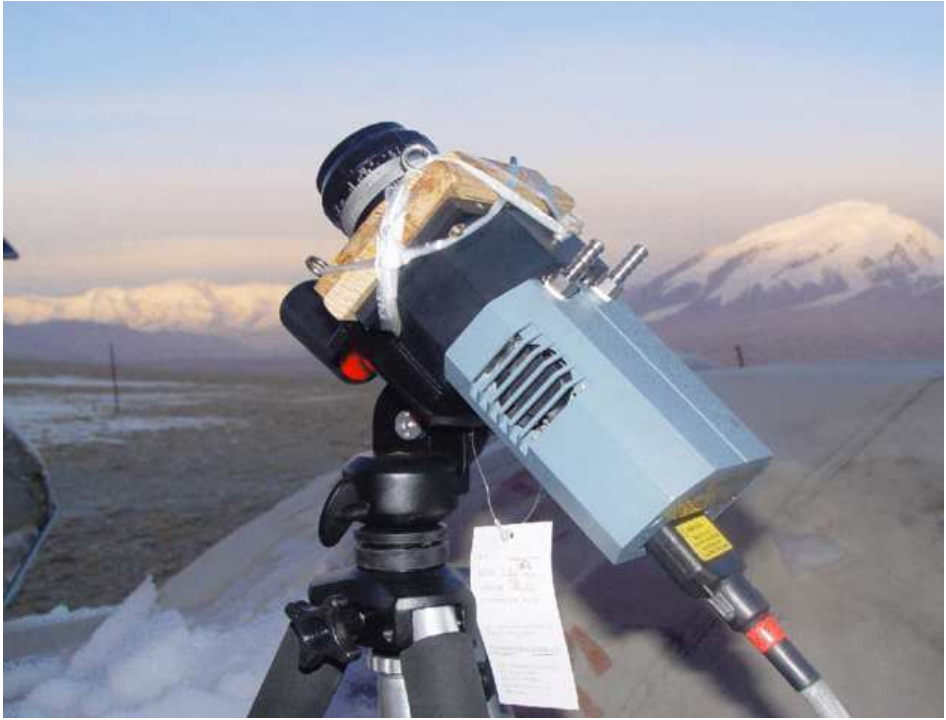}
\caption{The Andor CCD enclosed in its control box. This picture
was taken at Kalasu in the Tajik Autonomous County of Taxkorgan,
Xinjiang Pamirs of China.}\label{fig5}
\end{figure}

An Andor DV435 1K$\times$1K frame transfer CCD with a pixel size of 13
$\mu$m is used for the detector. Frame transfer technology is ideal
for fast imaging as it has the advantage of requiring no mechanical
shutter.  Avoiding the need for moving parts is very desirable on the
Antarctic plateau. The CCD was enclosed in a control box, as shown in
Fig.\,\ref{fig5}. The cable at the back of the box connects to the PCI
controller card installed in the control computer. The typical readout
noise of the CCD is $\sim 3\,e$ with maximum of $\sim 12\,e$, and the
gain is set to 2.0 $e$ per A/D. The peak quantum efficiency of the
Andor CCD at $-90\ssd$ is $\sim95$\%. During the typical exposure time
of 30\,s and under the typical ambient temperatures of less than
$-50\ssd$ on the Antarctic plateau, the dark current of the Andor CCD
is only $0.5\,e$. The dark current  can thus be negligible under
Antarctic conditions.

\begin{figure}
\includegraphics[width=125mm]{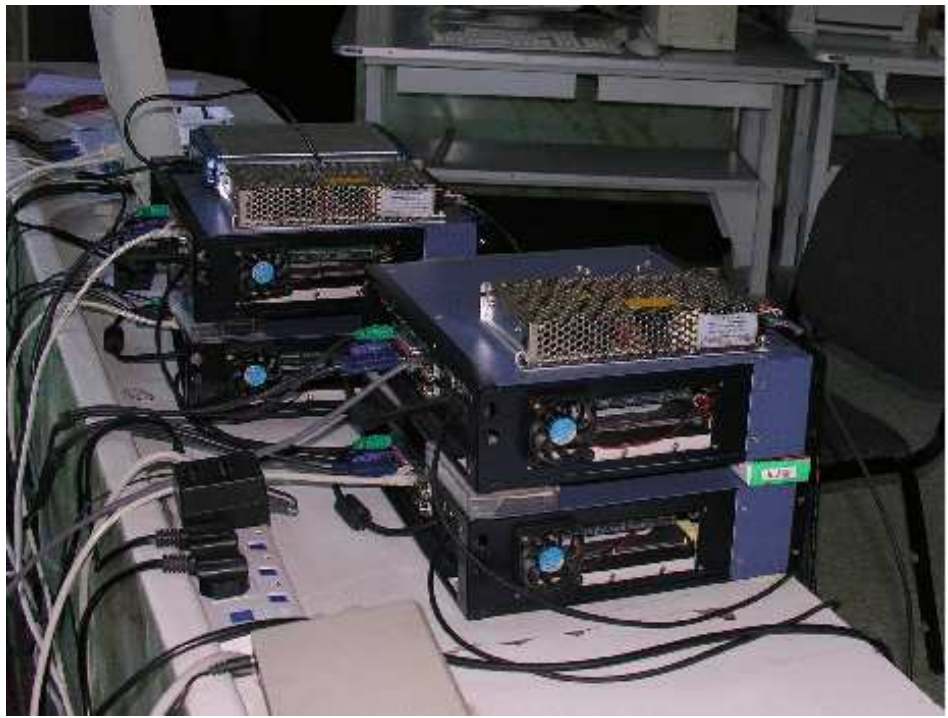}
\caption{Computer control equipment.}\label{fig6}
\end{figure}

Each Andor CCD is controlled through the CCI-010 PCI controller card
installed in an industrial control computer for each telescope. The
control computer is composed of a 1TX-i7415VL main board, Intel
Centrino 1.6 GHz CPU, and 1 GB of memory. Two kinds of storage disks
are used for the control computer. One is a 4 GB CompactFlash (CF)
disk which can operate at low temperatures (down to nearly $-45\ssd$);
the other is a normal 750 GB IDE hard disk. Fig.\,\ref{fig6} shows the
four control computers. Each computer weighs 8.3\,kg. The Windows
operating system is installed onto the CF disk because of its greater
reliability under low temperature conditions, while the 750 GB hard
disk is mainly used as data storage. The CCD control and data
collection software were developed based on the Andor-SDK-CCD software
development kit for the Windows-XP operation system. The time of the 
controler computer of \cstar$\#3$ is synchronized by GPS and the other 
computer correct its clock by \cstar$\#3$.

The real time data reduction process start automatically after the 
controller computer booting. The image coorected by bias and 
flat-field frames, and the catalogues objects is produced. The brightest
3000 stars of the catalogue from 1/3 images is moved to a special
directory for data transfere via iridium satellite communication.

\section{testing and data reduction}\label{obs}
\subsection{Testing}

In order to assure the performance of \cstar under the extremely low
temperature conditions of Dome A, the CCD system and several different
industrial control computers were tested. Finally, the whole \cstar
system was tested at low temperature in the laboratory of NAOC. These
tests indicated that the four telescopes and the CCD can work at low
temperatures down to nearly $-80\ssd$, while the four control
computers can work down to $-30\ssd$. In 2007 February 6 -- 9, the CCD
and control computers were tested at Kalasu. Kalasu (see
Fig\,\ref{fig5}) is located in the Tajik Autonomous County of
Taxkorgan, on the Xinjiang Pamirs of China at an elevation of 4450
m. We chose Kalasu as the test site because of its low temperature and
low atmospheric pressure conditions similar to the Antarctic
plateau. The atmospheric pressure was $\sim 58.6$ kPa and the
temperatures ranged from $-5\ssd$ to $-18\ssd$ during the testing
process. Both the CCD and the control computers were shown to work
normally during the two days of testing, and there are 4 750GB normal
hard disks were selected as data storage of CSTAR.

\begin{figure}
\includegraphics[width=85mm,height=85mm]{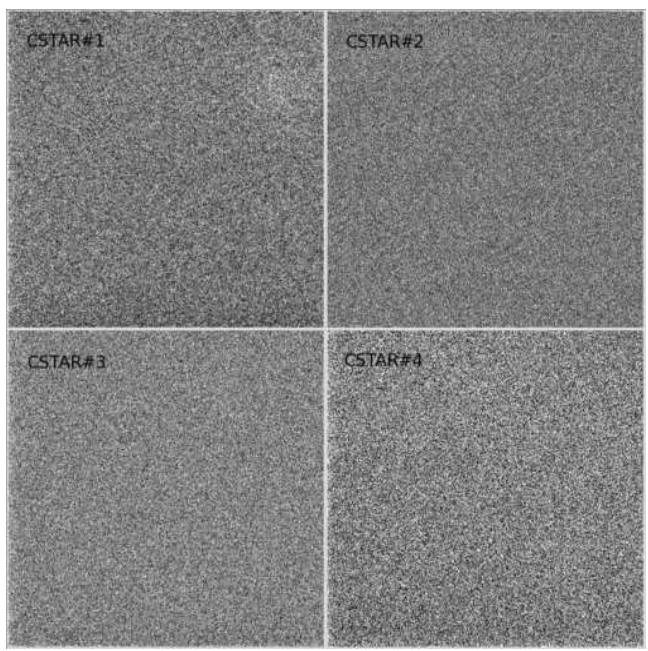}
\caption{Bias frame images for each telescope.}\label{bias}
\end{figure}

In 2007 September 3 -- 7, test observations of \cstar were performed
at the Xinglong station of NAOC. The four CCDs were cooled down to
$-40$ -- $-50 \ssd$ by electronic cooling system of the camera.  
The weather was good in most of  the time during
four observation nights, and more than 20,000 images were obtained.
The typical exposure time was 20\,s. Fig\,\ref{bias} shows the `super'
bias images for each telescope, which are the median of 100 bias frame
images for each telescope. There is no obvious variation and structure
in the `super' bias images. These `super' bias images are unique bias
frames to be used for reduction of data both from observations at
Xinglong and also from Dome A.

\begin{figure}
\includegraphics[width=85mm]{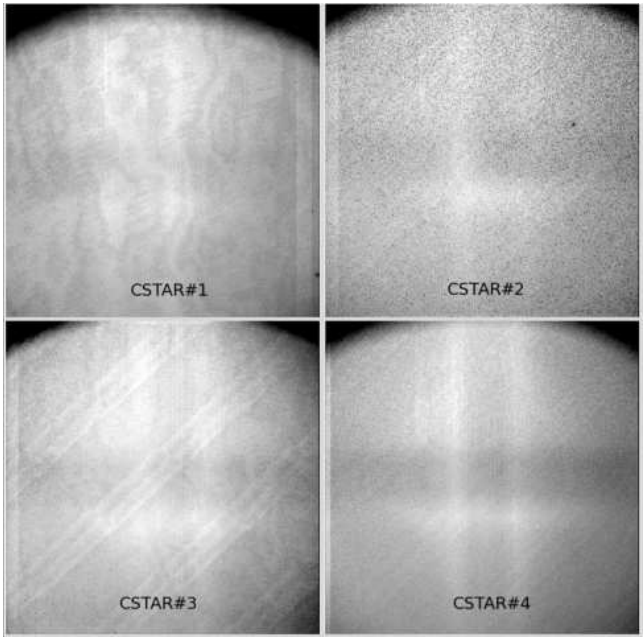}
\caption{Flat-field images for each telescope.}\label{flat}
\end{figure}

Variations of night-sky background are obvious even in the zenith
direction. If one takes the time during a photometric, moonless
night to obtain a long series of sky-dominated images pointing
directly at the zenith, the effects of the nonuniformity of the
night sky can be minimized. However, our telescope observes the
polar sky area at an airmass of 1.54 at Xinglong station. The median
sky background can only be used as an initial flat-field for
image correction. Thus, we typically obtained `supersky' flat-fields
by combining images of the sky \citep{zh04}. During this combination,
the bright stars in the images were masked and rejected, and only the
areas free from stars were used. By comparing the images, the
median level of each pixel could be selected to derive the final `supersky'
flat-field. 100 images of `supersky' for each telescope of \cstar
were used to obtain the `supersky' flat-field. These flat-fields
mostly reflect the small, pixel to pixel variations in the images.
Fig.\,\ref{flat} shows the final `supersky' flat-field images for
each telescope. Some obvious structures can still be seen.

\subsection{Data reduction}
First, for each filter a `super' bias frame was subtracted from each
image, then the `supersky' flat-field was divided by the
bias-corrected images. The bias and flat-field corrected data of $\sim
20000$ images obtained by \cstar during the four test-observation
nights were processed with the automatic data reduction software
developed by Z. J. JIANG and X. ZHOU based on the DAOPHOT photometric
package \citep{dao}, which was used in the data reduction of BATC
\citep{f96,wu07}. Because \cstar has a large field and is
undersampled, obtaining an accurate point-spread function (PSF) for
the sources detected across the whole view of field is very
difficult. The DAOFIND program was used to find stars in each image
and DAOPHOT was used to perform synthetic aperture photometry on the
objects detected by DAOFIND. All instrumental magnitudes of the four
telescopes were then normalized to the $V$ band magnitudes of stars in
the image 39530013.fit, which was observed by $\#3$ telescope on 2007
September 5.

\subsection{Error analysis and correction}
There are obvious systematic errors in the derived aperture-photometry
magnitudes. The errors mainly come from following sources:
\begin{enumerate}
\item The bias stability of each CCD\\ Due to the continuous
  observation during exposures and the frame transfer mode of the CCD,
  there is no opportunity to obtain real-time bias frames. The bias
  frames obtained at one time must be used for observations from
  another day at Dome A. Because of variations in the environmental
  parameters, such as temperature and instrumental status, the bias of
  each CCD camera may change. This variable bias will introduce linear
  errors in the observed magnitudes.
\item Non-uniformity of the `supersky' flat-field.\\ The flat-field
  images were not obtained during ideal photometric nights, and not
  from the zenith sky. A brightness gradient and asymmetry may exist
  in the flat-field frames. The variation in temperature from $-40$ to
  $-80\ssd$ may also change the characteristics of the flat
  field. During the polar observations by the fixed \cstar telescopes,
  every star will trace out a circle on the CCD, and the residual
  flat-field error will give a false variation in the observed
  magnitude of each star.
\item Variable PSF for stars in different positions in the images of
  \cstar.\\ The telescopes of \cstar have a large field of view. The
  optical design cannot keep the PSF exactly uniform over all parts of
  the image. When we use a fixed aperture to measure the magnitudes of
  the stars, the PSF depends on the location on the image and this
  will cause a variation in the instrumental magnitudes of each star
  relate to the other stars.
\end{enumerate}

\begin{figure}
\includegraphics[width=85mm]{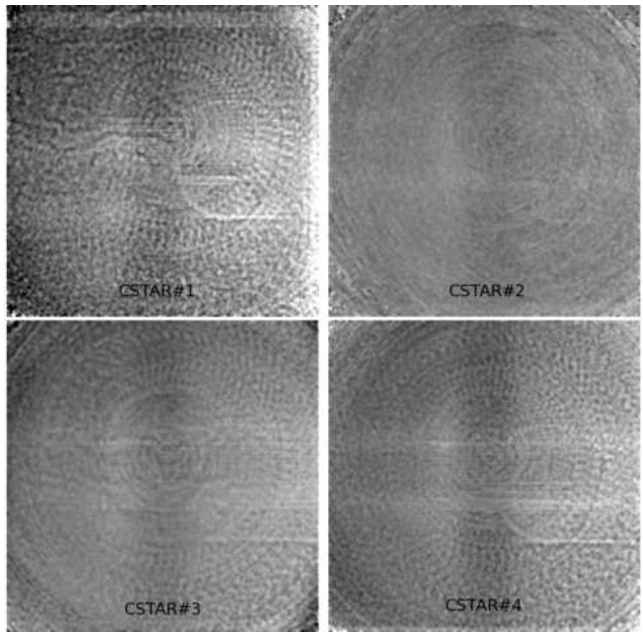}
\caption{Images of the corrected flat field for each telescope.}\label{mc-flat}
\end{figure}

Because we are observing a single area of the sky, and the sky's image
is rotating on the CCD, we have the opportunity to  correct the main
residual system errors mentioned above. Using thousands of stars with
very different magnitudes, we can easily determine the variable
component of the bias residuals based on the different magnitudes of
those stars in two different images. Using all of the circular traces
of the stars, the large-scale residual flat-field correction can be
obtained. Using the instrumental magnitudes from several different
apertures for each star, the aperture photometry curve-of-growth can
be obtained in all parts of the image but is mainly corrected with
residual flat-field correction mentioned above. The instrumental
magnitude obtained from different aperture were calibrated to the
standard system. After all these corrections, the systematic errors in
the derived photometric magnitudes can be reduced to the level of 0.01
mag for the brightest stars in most of the images. Some sudden
abnormal variations, where they exist, mostly come from the cirrus
clouds in the sky.  Fig.\,\ref{mc-flat} presents the
magnitude-corrected flat-field images for each telescope using
thousands of stars., and shows the obvious circular structures that
match the traces of stars on the CCD.

\begin{figure}
\includegraphics[width=125mm,angle=-90]{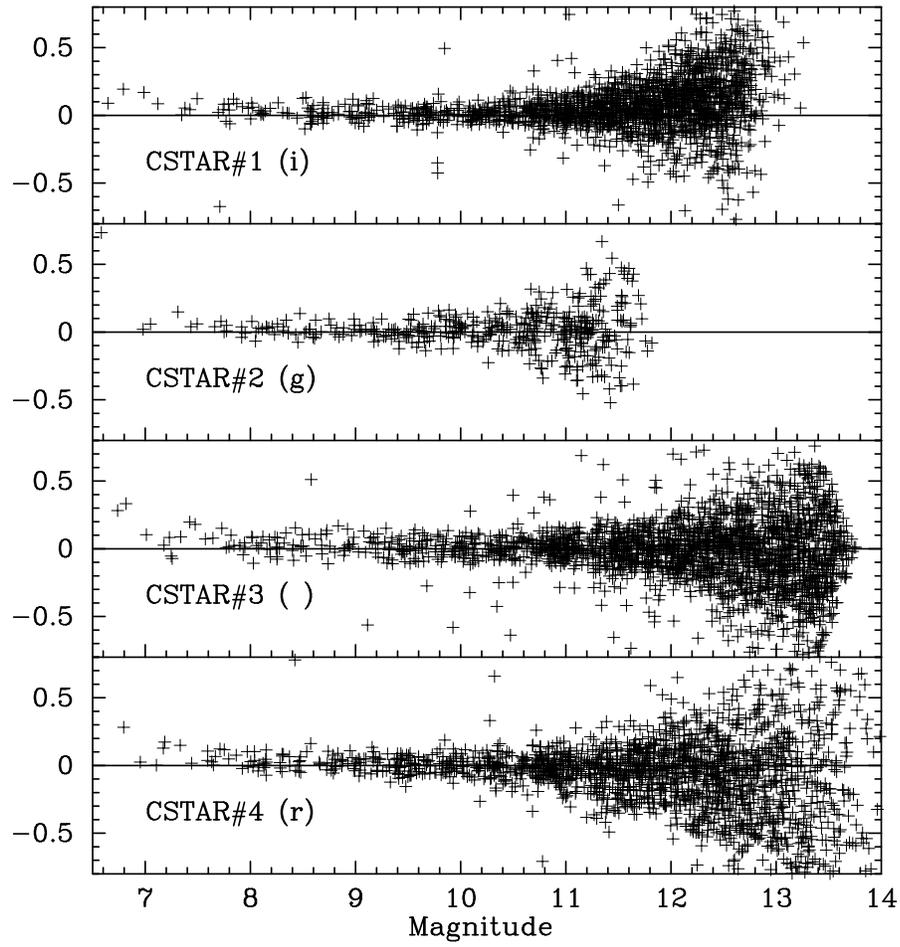}
\caption{Photometric errors for each telescope of \cstar. The vertical scale is in magnitudes.}\label{em}
\end{figure}

Two kinds of error estimates have been performed. One is theoretic 
statistical estimation based on star's magnitude and its sky
background. The other is obtained by real repeated observation of all the
objects in the images. By comparing the errors resulting from different 
images of
the same field with the same filter, we find that the measurement
errors are normally $\pm 0.01$ mag for bright stars. The statistical
errors can be regarded as the lower limits of the  measurement
errors. In the error estimates, we ignore points with abnormally large
deviations to calculate the root mean square (rms) errors. The abnormal 
variation may come from the true star brightness variation, or defect of 
the image (cosmic ray, satellites, bad pixels, etc). Fig.\,
\ref{em} shows the photometric errors for each telescope of \cstar
at different magnitudes. Because the errors estimated by this method
include real statistical errors and residual system errors from the bias and
flat-field correction, the errors shown in Fig.\,\ref{em} should be
larger than the actual  observational errors.
Fig.\,\ref{em} also shows that the efficiency of telescope \#2 of
\cstar is very low and that the limiting magnitude of this telescope is
about 2 mag lower than that of the other three telescopes. We knew
that the CCD camera for CSTAR \#2 was much noisier than the others, but we
were unable to change it in the time available.

\begin{figure}
\includegraphics[width=125mm]{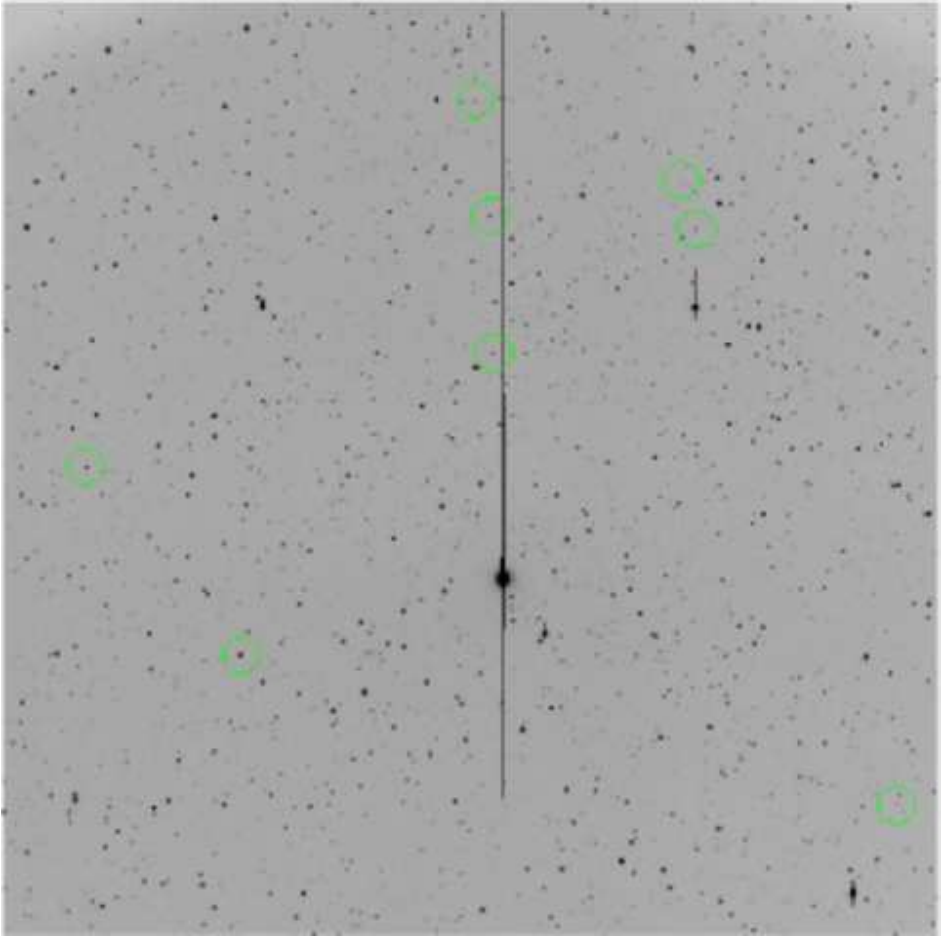}
\caption{The image 39530013.fit obtained by \cstar. The variable stars detected by
\cstar are labelled by green circles in the image.}\label{chart-var}
\end{figure}

\begin{figure}
\includegraphics[width=150mm]{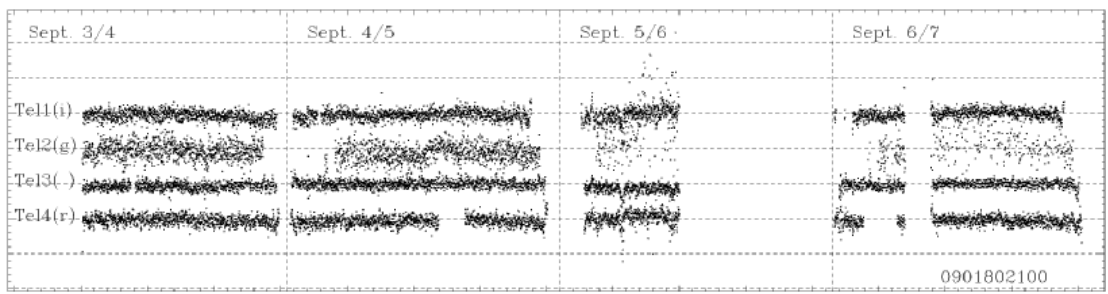}
\caption{The light curve of one of the bright sources from the \cstar telescopes.}\label{A0901802100}
\end{figure}

\begin{figure}
   \centering
   \includegraphics[width=30mm,height=\textwidth,angle=-90]{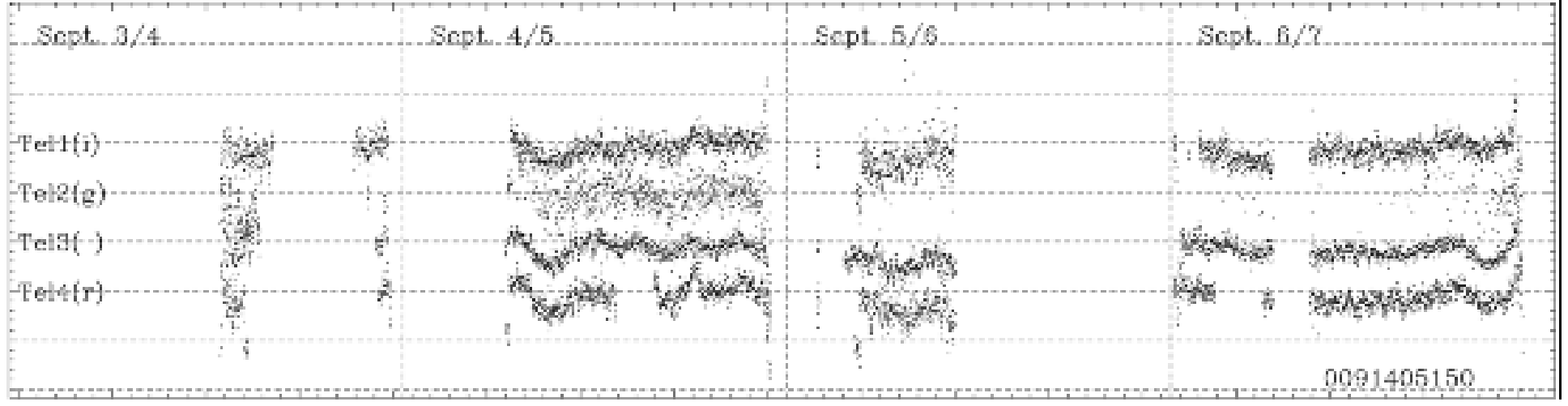}
   \includegraphics[width=30mm,height=\textwidth,angle=-90]{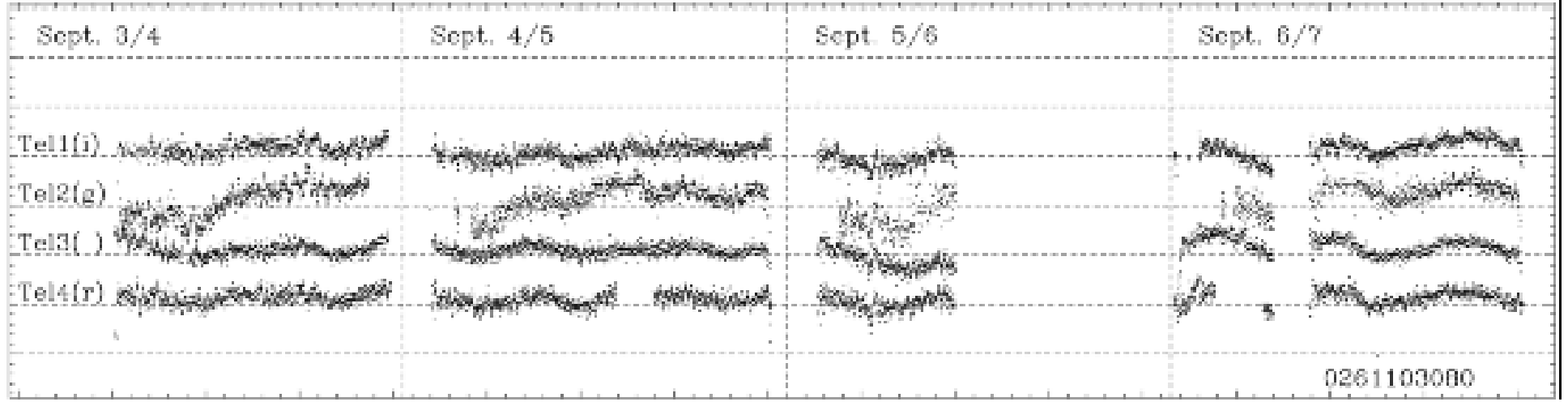}
   \includegraphics[width=30mm,height=\textwidth,angle=-90]{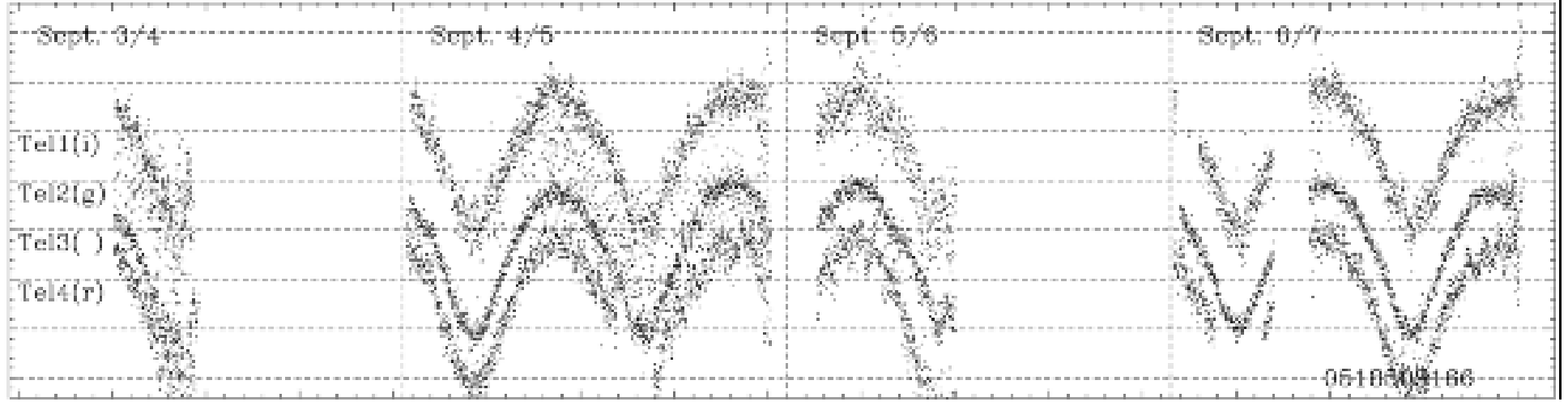}
   \includegraphics[width=30mm,height=\textwidth,angle=-90]{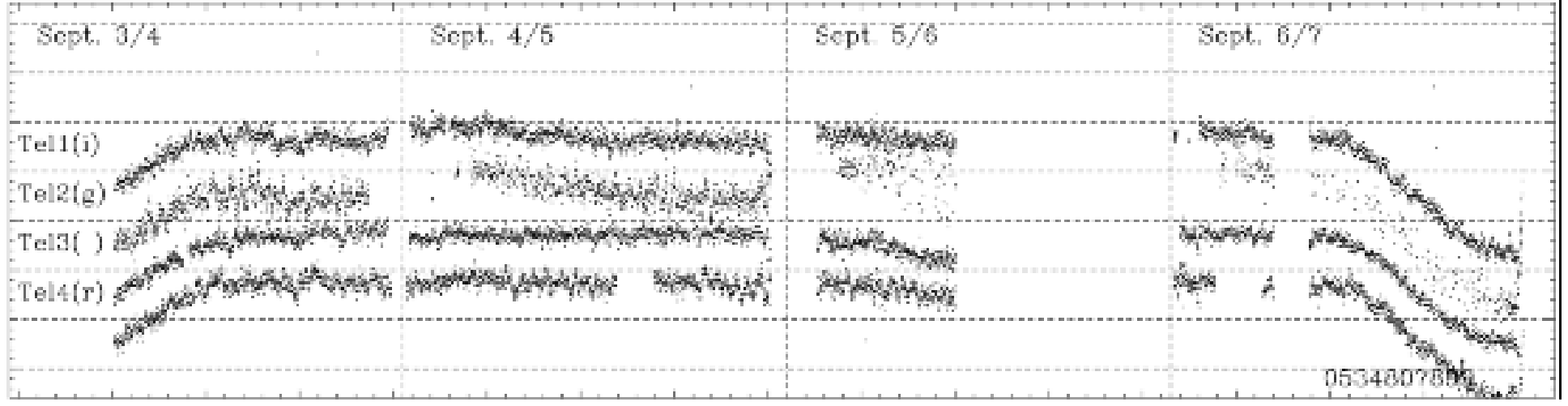}
   \includegraphics[width=30mm,height=\textwidth,angle=-90]{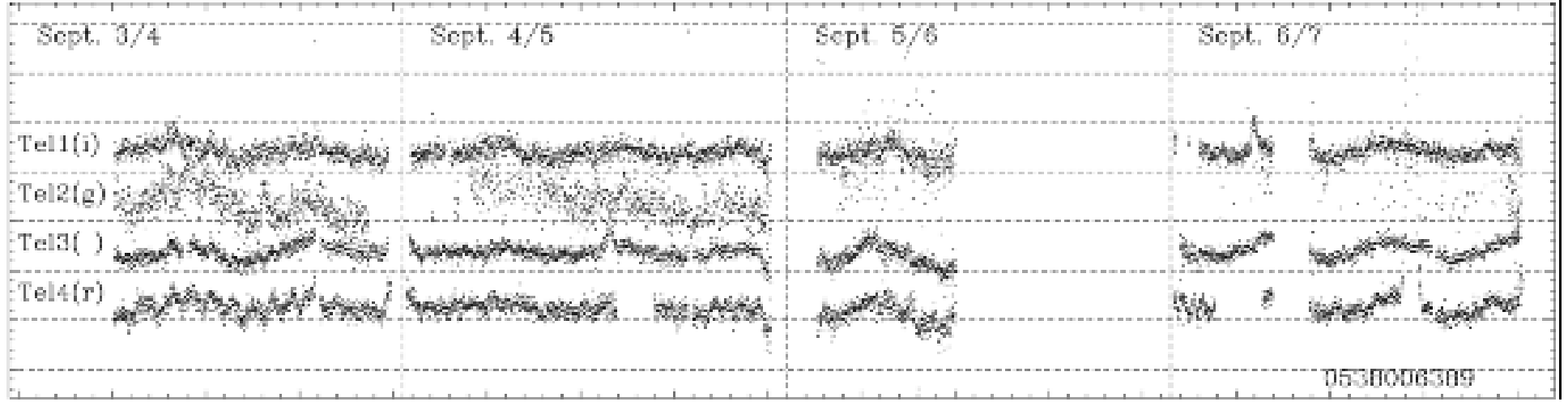}
   \includegraphics[width=30mm,height=\textwidth,angle=-90]{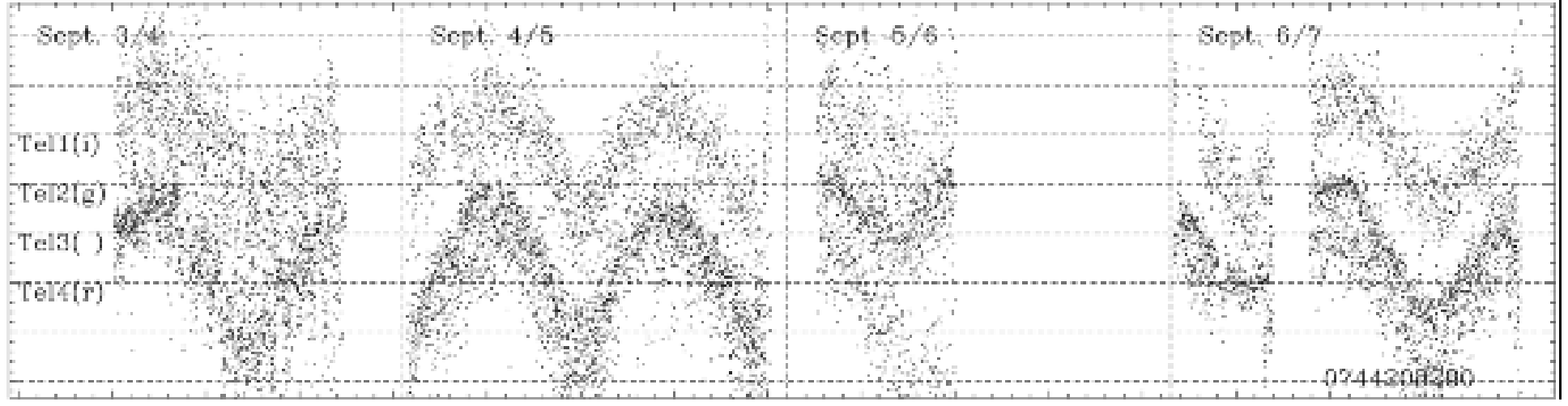}
   \includegraphics[width=30mm,height=\textwidth,angle=-90]{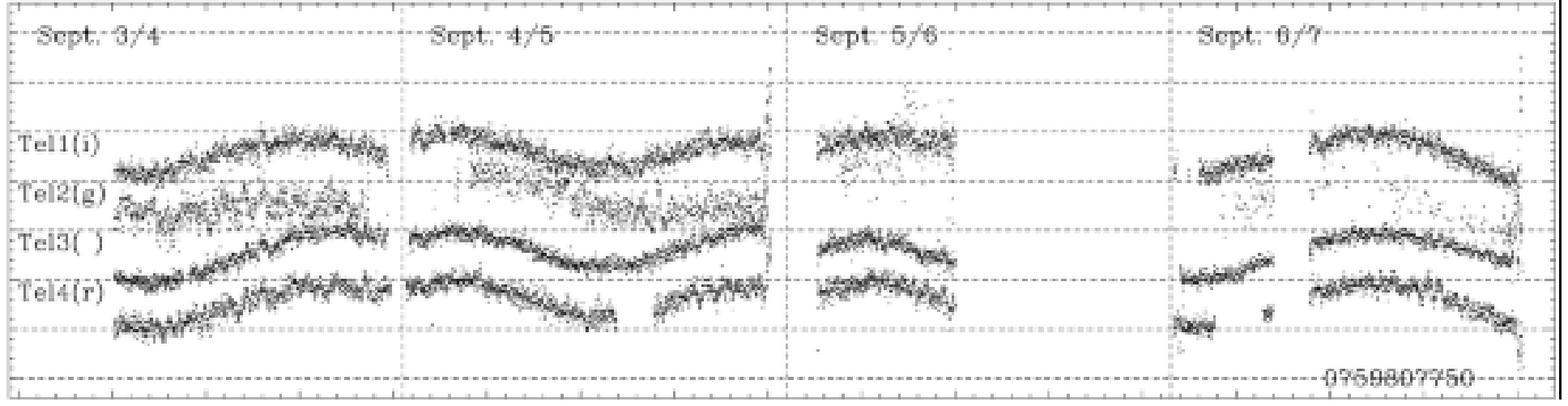}
   \includegraphics[width=30mm,height=\textwidth,angle=-90]{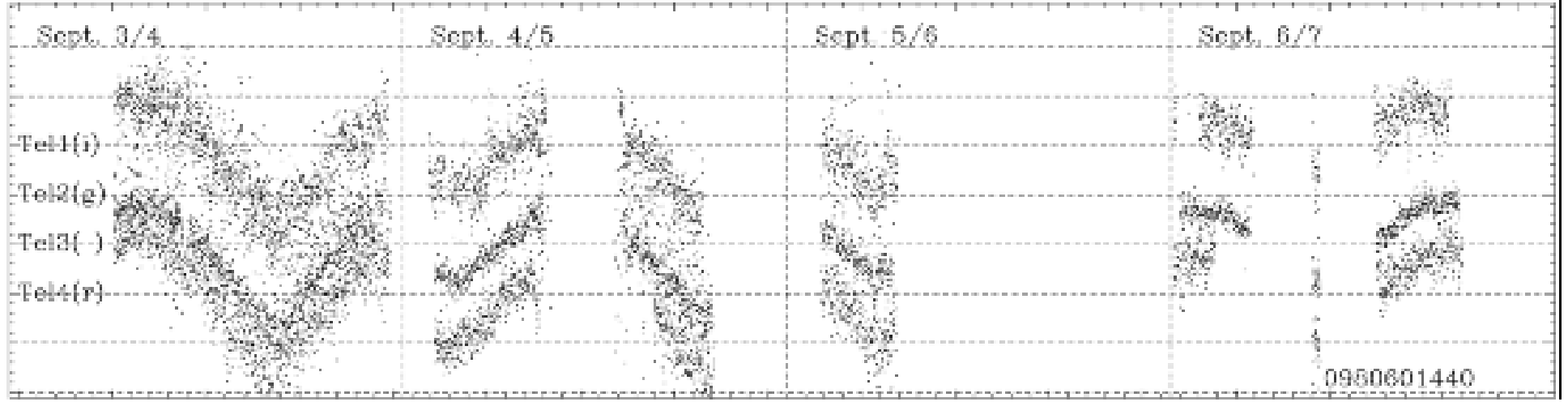}
     \caption{Light curves of  variable stars detected by \cstar during the test observations at the Xinglong station of NAOC. }
     \label{var}
\end{figure}

As an example of the data obtained, the light curve of one of the
bright stars from the four \cstar telescopes are shown in
Fig.\,\ref{A0901802100}.  The main scientific objectives of \cstar are
to assess the site quality of Dome A and to study the variable objects
in the region of the South
Pole. Fig.\,\ref{chart-var} shows the image 39530013.fit obtained by
\cstar during the test observations at the Xinglong station of
NAOC. The variable stars detected by \cstar are labeled by green
circles. The light curves of those variable stars are presented in
Fig.\,\ref{var}.

\section{conclusions}\label{con}
CSTAR, China's first Antarctic astronomical instrument is
described. \cstar is composed of four small Schmidt-Cassegrain
telescopes. Each telescope has an effective aperture of 100 mm and a
field of view of $\sim 4.5\dg\times4.5\dg$. Three of the four
telescopes are equipped with $g$, $r$, $i$ filters, the fourth one is
filter-less. A frame-transfer Andor DV435 1K$\times$1K CCD is used as
the detector on each telescope. A specially designed control computer
for each telescope is used for data acquisition and data reduction.

Low-temperature laboratory testing demonstrates that the telescopes
and the CCD can work under extremely low temperature (down to nearly
$-80\ssd$), while the control computer can work at temperatures as low
as $-30\ssd$. Actual test observations at Kalasu in the Xinjiang
Pamirs indicated that the CCD and control computer can work at these
low temperatures and under low atmospheric pressure conditions.

`Super' bias and `supersky' flat-field images were obtained during the
test observations at the Xinglong station of NAOC. These test
observations and the subsequent data reduction indicate that \cstar
can work stably and obtain a large volume of scientific data. A
special data reduction method was used to reduce the observational
errors for each of the objects detected by \cstar. The data reduction process
is done automatically in real time, and catalogue of brightest star from 
1/3 of the images obtained are prepared for further data transfer 
via iridium satellite communication. Finally, Eight variable
stars were detected by \cstar during the test observations.

\begin{acknowledgements}
This work was supported by the Chinese National Natural Science Foundation grands No. 10873016,
10633020, 10603006, and 10803007, and by National Basic Research Program of China (973 Program),
No. 2007CB815403. We thank our colleagues at the University of New South Wales, Australia, for 
assistance in editing this paper.

\end{acknowledgements}

\begin{table}
\begin{minipage}[]{100mm}
\caption[]{Passband parameters of filters used by \cstar.\label{tb1} }
\end{minipage}
\begin{tabular}{ccccc}
\hline\noalign{\smallskip}
         telescope       & \cstar\,\#2 & \cstar\,\#4 & \cstar\,\#1 & \cstar\,\#3  \\
  \hline\noalign{\smallskip}
           filter        &   $g$    &  $r$      &  $i$   &   none      \\
 effective Wavelength (nm)   &  470  &  630  &  780   &   \\
 FWHM (nm)                    &  140  &  140  &  160   &  \\
  \noalign{\smallskip}\hline
\end{tabular}
\end{table}

\end{document}